\documentclass[12pt, a4paper]{article}
\pdfoutput=1

\usepackage{rotating}
\usepackage{amsmath}
\usepackage{mathtools}
\usepackage{amsthm}
\usepackage{amsfonts}
\usepackage[dvips]{epsfig}
\usepackage{graphicx}
\usepackage{psfrag}

\usepackage{amssymb}
\usepackage{lscape}
\usepackage{slashed}
\usepackage{lscape}
\usepackage{color}
\usepackage{cite}
\usepackage{url}
\usepackage{caption}
\usepackage{bm}
\usepackage{subfig}
\usepackage{caption}

\newcommand{\beq}{\begin{equation}}
\newcommand{\eeq}{\end{equation}}
\newcommand{\ga}{\lower.7ex\hbox{$\;\stackrel{\textstyle>}{\sim}\;$}}
\newcommand{\la}{\lower.7ex\hbox{$\;\stackrel{\textstyle<}{\sim}\;$}}

\makeatletter
\newcommand{\Cen}[2]{%
  \ifmeasuring@
    #2%
  \else
    \makebox[\ifcase\expandafter #1\maxcolumn@widths\fi]{$\displaystyle#2$}%
  \fi
}
\makeatother

\begin{document}

\begin{flushright}
{\tt LPT--Orsay 17-36}
{\tt UMN--TH--3635/17}
{\tt FTPI--MINN--17/15}
\end{flushright}

\vspace{0.5cm}
\begin{center}
{\bf {\large Enhancement of the Dark Matter Abundance Before Reheating:
Applications to Gravitino Dark Matter}
}
\end{center}

\vspace{0.05in}

\begin{center}{
{\bf Marcos~A.~G.~Garcia}$^{a}$,
{\bf Yann Mambrini}$^{b}$},
{\bf Keith~A.~Olive}$^{c,d}$, and
{\bf Marco Peloso}$^c$
\end{center}

\begin{center}
 {\em $^a$Physics and Astronomy Department, Rice University, Houston, TX
 77005, USA}\\[0.2cm] 
 {\em $^b$Laboratoire de Physique Th\'eorique 
Universit\'e Paris-Sud, F-91405 Orsay, France}\\[0.2cm]
  {\em $^c$School of
 Physics and Astronomy, University of Minnesota, Minneapolis, MN 55455,
 USA}\\[0.2cm] 
{\em $^d$William I. Fine Theoretical Physics Institute, School of
 Physics and Astronomy, University of Minnesota, Minneapolis, MN 55455,
 USA}

\end{center}

\bigskip

\centerline{\bf ABSTRACT}

\noindent  
In the first stages of inflationary reheating, the temperature of the radiation produced by inflaton decays is typically
 higher than the commonly defined reheating temperature $T_{RH} \sim (\Gamma_\phi M_P)^{1/2}$ where $\Gamma_\phi$
is the inflaton decay rate. We consider the effect
of particle production at temperatures at or near the maximum temperature attained during reheating. We show that the impact of this early production on the final particle abundance
depends strongly on the temperature 
dependence of the production cross section. For $\langle \sigma v \rangle \sim T^n/M^{n+2}$, and for $n < 6$, any particle 
produced at $T_{\rm max}$ is diluted by the later generation of entropy near $T_{RH}$. This applies to
cases such as gravitino production in low scale supersymmetric models ($n=0$) or NETDM models of dark matter ($n=2$).
However, for $n\ge6$ the net abundance of particles produced during reheating is enhanced by over an order of 
magnitude, dominating over the dilution effect. This applies, for instance to gravitino production in high scale supersymmetry models where $n=6$.

\vspace{0.2in}

\begin{flushleft}
August 2017
\end{flushleft}
\medskip
\noindent

\newpage

\section{Introduction}

One of the key attributes of inflationary cosmology \cite{reviews} is its independence of initial conditions.
Once inflation commences, all prior history is inflated away, and the universe 
begins afresh with new nearly homogeneous and isotropic initial conditions which
depend primarily on the reheating process after inflation. In its simplest form, reheating occurs
as the inflaton settles to its minimum after inflation and the coherent scalar field oscillations of the
inflaton decay. If the decay products thermalize rapidly, a radiation temperature is established,
and in the limit of instantaneous decay and reheating at $\Gamma_\phi \sim H$ where 
$\Gamma_\phi$ is the inflaton decay rate and $H$ is the Hubble parameter, we can define a reheating
temperature as $T_{RH} \sim (\Gamma_\phi M_P)^{1/2}$, where $M_P = (8 \pi G_N)^{-1/2}$ is the reduced Planck mass \cite{dg,nos}.

In reality, inflaton decay is not instantaneous, though thermalization may indeed be quite rapid \cite{therm,EGNOP}.
If thermalization is rapid, then the early inflaton decay products can achieve temperatures significantly higher than $T_{RH}$ \cite{Tmax1,Tmax2,EGNOP}.
In turn, this may significantly alter the production rate and abundance of particles which are weakly coupled to the thermal bath.
The gravitino is a prime example. Gravitinos are produced during reheating and their abundance is typically proportional to the reheating temperature \cite{nos,ehnos,kl,ekn,Juszkiewicz:gg,mmy,Kawasaki:1994af,Moroi:1995fs,enor,Giudice:1999am,bbb,Pradler:2006qh,ps2,rs,EGNOP}. Although the rate for gravitino production is enhanced at temperatures
above $T_{RH}$, gravitinos produced at $T > T_{RH}$ are diluted by the bulk of the entropy produced in subsequent inflaton decays. 
These (non)-results are, however, specific to the cross sections that characterize the particle production. 

Here we consider particle production during reheating at temperatures $T > T_{RH}$. We consider a general form for the temperature 
dependence of the production cross section. We then apply these results to three specific cases. 
1) The gravitino, as discussed above in models of low energy supersymmetry.
2) Non-equilibrium thermal dark matter \cite{moqz} models. These are models where the dark matter candidate
couples to the thermal bath through the exchange of some massive mediator. As a result, they never attain
thermal equilibrium, yet are produced from the thermal bath. While similar to the gravitino, the details of the production
mechanism differs. 3) We return to gravitinos in the case of high scale supersymmetry, where all superpartners (other than the gravitino)
have masses above the inflaton mass \cite{bcdm,DMO}. In this case, gravitinos can not be singly produced but rather
can only be produced in pairs. Once again, the details of the production mechanism differs from the previous two cases.

The paper is organized as follows. In section 2, we write down the relevant equations for generalized particle production and describe the three specific models we use as examples. In section 3, we derive the abundance of particles produced assuming instantaneous reheating and 
derive the effect of particle production at $T> T_{RH}$ in section 4. In section 4, we also provide some numerical
results to support the analytic approximations made. 
Our conclusions are given in section 5.

\section{Dark matter production at reheating}

For our analysis, we first need to compute the dark matter production at early stages of reheating. 
We can define the thermally averaged cross section 
\beq\label{sigmav}
\langle \sigma|v|\rangle = \frac{ \lambda T^{n}}{\pi M^{n+2}}\,,
\eeq
for dark matter production, where we assumed a dark matter mass $m_\chi \ll T$, and that $\chi$ is coupled to the thermal bath by a heavy mediator of mass $m_X \gg T$. In this case, the mass scale $M$ in (\ref{sigmav}) is parametrically related to the mediator mass, $M \sim m_X$. For the case of the gravitino, one should associate the 
scale $M$ with the supersymmetry breaking scale, $F$ which may be related to the geometric  mean of the Planck scale $M_P$, and gravitino mass, $m_{3/2}$ for the production of longitudinal modes of the gravitino.

Reheating is a finite duration process that starts at the end of inflation, and is concluded with the formation of a dominant thermal bath due to inflaton decay. Assuming instantaneous thermalization of the inflaton decay products \cite{therm,EGNOP}, this thermal bath reaches the maximum temperature $T_{\rm max}$ shortly after inflation ends, when only a small fraction of the inflaton energy has decayed, and the energy density of the universe is still dominated by the inflaton mass. This temperature may be orders of magnitude greater than the reheating temperature $T_{RH}$, that is achieved later on, when most of the inflaton energy has decayed, and the thermal bath has become dominant \cite{EGNOP, Tmax1,Tmax2}. Most computations of relic abundances from the early universe assume an instantaneous inflaton decay into a thermal bath of temperature $T_{RH}$. These computations ignore any production that took place during reheating (namely, while the thermal bath was subdominant, as its temperature decreased from $T_{\rm max}$ to $T_{RH}$). This approach is valid as long as the production rate in eq.~(\ref{sigmav}) is $not$ competitive with the dilution rate due to the inflaton decay, which is (as we will demonstrate) not always justified. 

In this section, we propose to precisely quantify the validity of this assumption, by comparing the dark matter production obtained supposing an instantaneous reheating (subsection  \ref{subsec:inst-reh}) with the complete process, that accounts for the finite-time duration of the inflaton decay (subsection \ref{subsec:inst-therma}). We will see that the degree of accuracy depends on the specific value of the exponent $n$ in the temperature dependence $T^n$ of the thermally averaged cross-section (\ref{sigmav}). We will then discuss three different microscopic/UV models, characterized by three different values of $n$.

\subsection{Instantaneous reheating} \label{subsec:inst-reh}

Under the assumption of instantaneous reheating, the inflaton instantaneously decays into a thermal bath of initial temperature 
\cite{dg,nos}
\beq\label{trehdef}
T_{RH} = \left(\frac{40}{g_{RH}\pi^2}\right)^{1/4}\left(\frac{\Gamma_{\phi}M_P}{c}\right)^{1/2}\,, 
\eeq
which dominates the energy density of the universe, where $\Gamma_{\phi}$ is the inflaton decay rate, $g_{RH} \equiv g \left( T_{RH} \right)$ is the number of effective degrees of freedom in the thermal bath, and  $c$ is an order one parameter that depends on when exactly the decay is assumed to take place. For instance, $c=1$ if we set the decay time $t_{RH}=\Gamma_{\phi}^{-1}$, or $c=2/3$ if we set the Hubble rate $H(T_{RH})=\Gamma_{\phi}$. Numerical solutions to reheating give $c \approx 1.2$ \cite{ps2,EGNOP}. 
 In what follows we will set $c=1$ for definiteness.
 
Consider for instance the process $\gamma_1+\gamma_2\rightarrow \chi_1+\chi_2$, where $\gamma_{1,2}$ are constituents of the thermal plasma, and $\chi_{1,2}$ denote the scattering products, out of which $\chi_1$ or both $\chi_{1,2}$ correspond to the dark matter particle; in this section we assume for simplicity that $\chi_1=\chi_2\equiv \chi$. If the scattering cross section is small enough to keep the dark matter number density, $n_{\chi}$, well below its thermal equilibrium value, $n_{\chi}^{\rm eq}$, at all times, then the Boltzmann equation controlling the dark matter abundance $Y_\chi \left( T \right) \equiv \frac{n_\chi \left( T \right)}{n_{\rm rad} \left( T \right)}$ is of the form~\footnote{Here for convenience, $n_{\rm rad}$ is defined as the number density of a single bosonic relativistic degree of freedom in thermal equilibrium, $n_{\rm rad}=\zeta(3)T^3/\pi^2$. 
The final abundance $Y_\chi$ can be immediately related to the dark matter relic fractional density through 
$\Omega_{\chi} = \frac{m_{\chi} n_\chi}{\rho_c} = \frac{m_\chi \, Y_\chi n_{\rm rad}}{\rho_c}$, 
where $\rho_c$ is the critical energy density of the universe. }
%
\beq\label{Yequation}
\dot{Y}_{\chi} + 3\left(H + \frac{\dot{T}}{T}\right) Y_{\chi} = g_{\chi}^2\langle \sigma|v|\rangle n_{\rm rad}\,,
\eeq
where $H$ is the Hubble rate and $g_\chi$ is the number of degrees of 
freedom of $\chi$ (times 3/4 if $\chi$ is a fermion). This is solved by 
\beq\label{ansol1}
Y_{\chi}(T)=Y_{\chi}(T_{RH})\,\frac{g(T)}{g_{RH}} - g(T)\int_{T_{RH}}^T  \frac{g_{\chi}^2\langle\sigma|v|\rangle n_{\rm rad}(\tau)}{g(\tau)H(\tau)\,\tau}\left[1+\frac{\tau}{3} \frac{d\ln g(\tau)}{d\tau}\right]\,d\tau\,, 
\eeq
where  $g \left( T \right)$ is the number of effective relativistic degrees of freedom in the thermal bath at temperature $T$. 
We have assumed entropy conservation so that $g T^3 a^3 =$ const., where $a$ is the cosmological scale factor.   

We now use the thermal cross section (\ref{sigmav}), and assume a vanishing dark matter abundance at the beginning of reheating,  $Y_{\chi}(T_{RH})=0$. Accounting for the fact that  $g$ and the coupling $\lambda$ depend only weakly on the temperature, eq.~(\ref{ansol1}) integrates to
\begin{align}\notag
Y_{\chi,{\rm instant.}}(T) &\simeq - \frac{\zeta(3)\sqrt{90}\, g_{\chi}^2M_P}{\pi^4 M^{n+2}} g(T)\int_{T_{RH}}^T  \frac{ \lambda(\tau) \tau^n}{g(\tau)^{3/2}} \,d\tau \\ 
&\simeq  \frac{\zeta(3)\sqrt{90}\, g_{\chi}^2 M_P}{(n+1) \pi^4  M^{n+2}} g(T) \left[  \frac{\lambda(T_{RH}) T_{RH}^{n+1}}{g_{RH}^{3/2}} - \frac{\lambda(T) T^{n+1}}{g(T)^{3/2}} \right] \label{Ynaive}
\end{align}
which asymptotes to the value
\beq
Y_{\chi,{\rm instant.}} \simeq   \left(\frac{90}{g_{RH}}\right)^{1/2} \left(\frac{g(T)}{g_{RH}}\right)  \frac{\zeta(3) g_{\chi}^2 \lambda(T_{RH}) T_{RH}^{n+1} M_P}{(n+1) \pi^4 M^{n+2}}  \,, 
\label{Yinst}
\eeq
when $T < T_{RH}$. We have assumed
$n > -1$ in eq. (\ref{Yinst}) (so that the first term dominates in the square parenthesis of eq. (\ref{Ynaive})). 
We can then define 
\beq
R_{\chi,{\rm instant.}} (T) = \frac{Y_{\chi,{\rm instant.}}(T)}{Y_{\chi,{\rm instant.}}}
\label{Rinst}
\eeq
as the ratio of the temperature-dependent abundance relative to its asymptotic value.

In the next subsection we compare the result (\ref{Ynaive}), obtained under the assumption of instantaneous reheating, against the abundance obtained if we more properly account for the finite duration of reheating.

\subsection{Instantaneous thermalization} \label{subsec:inst-therma}

Reheating after inflation is a continuous process, that dumps the energy density of the inflaton into the relativistic plasma, while diluting the previously created content of the universe. Therefore, in order to track the relic dark matter density, one must solve the following system of equations
\begin{align}\label{set1}
& \dot{\rho}_{\phi} + 3H\rho_{\phi} + \Gamma_{\phi}\rho_{\phi} = 0\\ \label{set2}
& \dot{\rho}_{\gamma} + 4H\rho_{\gamma} - \Gamma_{\phi}\rho_{\phi} =0 \\ \label{set4}
& \dot{n}_{\chi} + 3Hn_{\chi} + \langle \sigma |v|\rangle \left[ n_{\chi}^2   - (n_{\chi}^{\rm eq})^2  \right] = 0 \\ 
& \rho_{\phi}+\rho_{\gamma}  = 3 \, M_P^2 \, H^2 \label{H2}
\end{align}
where $\rho_\phi$ and $\rho_\gamma$, are, respectively, the energy density of the inflaton and of the thermal bath formed by inflaton decay. We stress that we are assuming that the dark matter is not directly coupled to the inflaton, and it is only produced by the thermal bath with the cross section (\ref{sigmav}). We continue to assume instantaneous thermalization of the inflaton decay products, as justified in~\cite{therm,EGNOP}. Finally, we disregard the production of  dark matter in the second and fourth equations (\ref{set2} and \ref{H2}),
as we will work 
in the limit of small dark matter production, so that $\rho_\gamma$ and $H$ are negligibly modified by dark matter production. 

Solving the first two equations of the system one finds that the thermal bath reaches a maximum temperature $T_{\rm max}$ when only a small amount of the inflaton energy  has decayed \cite{Tmax1,Tmax2,EGNOP}. This temperature is much higher than the reheating temperature, defined to be the temperature of the thermal bath when it starts to dominate over the residual energy of the inflaton. One finds (see for instance  \cite{EGNOP}) 
\beq
T_{\rm max} \simeq 0.5\left(\frac{m_{\phi}}{\Gamma_{\phi}}\right)^{1/4}T_{RH}\,, 
\eeq
where $m_\phi$ is the inflaton mass. Perturbativity requires $\Gamma_{\phi} < m_\phi$, and it is not uncommon to have $\Gamma_\phi \ll m_\phi$ (this is for instance the case if the inflaton decays gravitationally). Therefore  $T_{\rm max}$ can be many orders of magnitude greater than $T_{RH}$, possibly leading to a larger production of dark matter.  This opens the question regarding the accuracy of the result (\ref{Ynaive}), that assumes that the temperature was never above  $T_{RH}$. On the other hand, most of the energy of the universe is still in the inflaton when $T = T_{\rm max}$, and the entropy generated by the subsequent decay of this energy dilutes the dark matter quanta produced at  $T \sim T_{\rm max}$. Given these two contrasting arguments, only an  explicit solution of the system (\ref{set1})-(\ref{set2})-(\ref{set4}) can shed light on the accuracy of the instantaneous reheating  result (\ref{Ynaive}). 
 
We assume that the inflaton performs coherent oscillations about the (quadratic) minimum of its potential at the end of inflaton. This leads to an equation of state for the inflaton $w=p/\rho=0$, when averaged over a complete oscillation (the oscillations occur on a timescale $m_\phi^{-1}$, which is much shorter than the other timescales of reheating, and taking $w=0$ for the inflaton is therefore a very accurate assumption). The inflaton dominates the energy density until the very end of reheating, so it is a good approximation to set $w=0$ for the whole duration of reheating. This will allow us to obtain an analytic result for the dark matter abundance, that we can compare with an exact numerical solution of the system  (\ref{set1})-(\ref{set2})-(\ref{set4}). Under this assumption, the scale factor evolves as~\cite{EGNOP}  
\beq\label{av}
\frac{a(t)}{a_{\rm end}} \simeq \left(1+\frac{v}{A}\right)^{2/3} \simeq \left(\frac{v}{A}\right)^{2/3}\,,
\eeq
with $v \equiv \Gamma_{\phi} \left( t-t_{\rm end}\right)$ (the suffix ``end'' indicating the end of inflation, when $w = -1/3$) and 
\beq
A \equiv \frac{\Gamma_{\phi}}{m}\left(\frac{3}{4}\frac{\rho_{\rm end}}{m^2M_P^2}\right)^{-1/2}\simeq \mathcal{O}(1)\, \frac{\Gamma_{\phi}}{m} \,,
\eeq
where the $\mathcal{O}(1)$ factor in the second equality is approximately equal to 2.8 for Starobinsky inflation, and to 1.3 for a quadratic potential. 

In the regime  $A\ll v\ll 1$, we obtain~\cite{EGNOP}   
\beq
\rho_{\gamma}\simeq \rho_{\rm end}A^2 v^{-8/3}\boldsymbol{\gamma}(5/3,v) \simeq \frac{3}{5}\rho_{\rm end}A^2 v^{-1} = \frac{4}{5}(\Gamma_{\phi}M_P)^2 v^{-1}\,,
\eeq 
where $\boldsymbol{\gamma}$  denotes the lower incomplete gamma function. This in turn,  implies 
\beq\label{Tv}
T\simeq \left(\frac{24}{\pi^2 g}\right)^{1/4}(\Gamma_{\phi}M_P)^{1/2}v^{-1/4}\,.
\eeq
With the scattering cross section given by (\ref{sigmav}), and $n_{\chi}^{\rm eq}= g_{\chi} n_{\rm rad}$, we can readily rewrite (\ref{set4}) as
\begin{align}\notag 
\frac{d}{dT}\left[n_{\chi}\left(\frac{a}{a_{\rm end}}\right)^3\right] &= \frac{g_{\chi}^2\langle \sigma|v|\rangle n_{\gamma}^2}{\dot{T}}\left(\frac{a}{a_{\rm end}}\right)^3\\ \label{diffeqn}
& = g_{\chi}^2\left(\frac{ \lambda T^{n}}{\pi M^{n+2}}\right)\left(\frac{\zeta(3)T^3}{\pi^2}\right)^2\left(-\frac{96 \Gamma_{\phi}M_P^2}{g\pi^2 T^5}\right) \left(\frac{24\Gamma_{\phi}^2M_P^2}{g\pi^2 T^4A}\right)^{2} \;, 
\end{align}
which is solved by  
\beq
n_{\chi}\left(\frac{a}{a_{\rm end}}\right)^3 = \frac{55296\zeta(3)^2 \, g_{\chi}^2\lambda\Gamma_{\phi}^5M_P^6}{g^3\pi^{11}M^{n+2}A^2} \times \begin{cases}
\dfrac{1}{n-6}\left(T_{\rm max}^{n-6} - T^{n-6} \right)\,, & n\neq 6\\[5pt]
\ln\left(\dfrac{T_{\rm max}}{T }\right)\,, & n=6
\end{cases}\,.
\eeq
Dividing this by $n_{\rm rad}$, we find, at the end of reheating,
\beq\label{Yabove}
Y_{\chi}^{(n)} (T_{RH}) = \frac{96\zeta(3)\, g_{\chi}^2\lambda M_P T_{RH}^{7}}{\sqrt{40} g_{RH}^{1/2}\pi^{4}M^{n+2}} \times 
\begin{cases}
\dfrac{1}{n-6}\left(T_{\rm max}^{n-6} - T_{RH}^{n-6} \right)\,, & n\neq 6\\[5pt]
\ln\left(\dfrac{T_{\rm max}}{T_{RH} }\right)\,, & n=6
\end{cases}\,.
\eeq

\begin{figure}[ht!]
\centering
    \scalebox{0.85}{\includegraphics{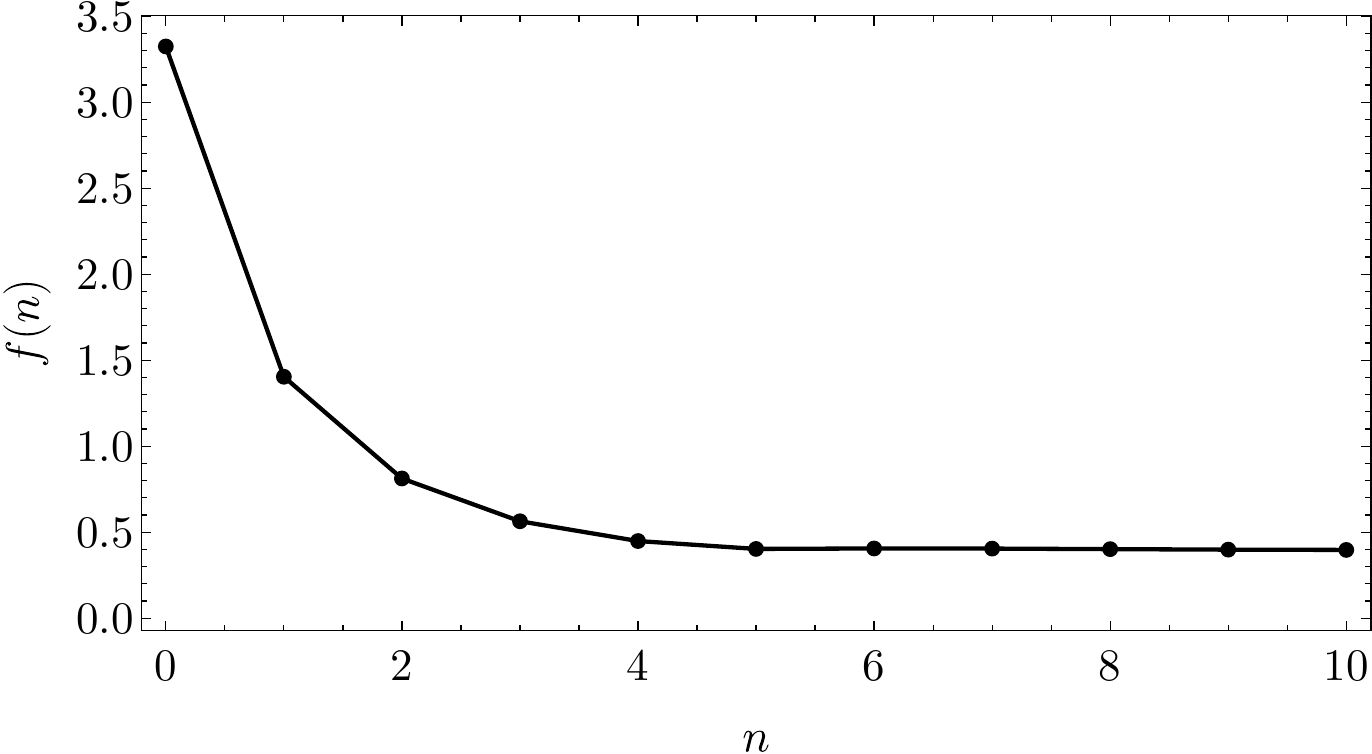}}
    \caption{\em Numerical correction to the analytical result~(\ref{Yratio}) for the ratio of the exact dark matter yield to the instantaneous approximation, $R_{\chi}^{(n)}$. The function $f(n)$ asymptotes to the value $\sim 0.4$ for large $n$ .
        }
    \label{fig:fnfig}
\end{figure}

We can now compare this result with (\ref{Yinst}),  obtained under the assumption of instantaneous reheating. At $T \ll T_{RH}$  we find 
\beq\label{Yratio}
R_{\chi}^{(n)}(T)\equiv \frac{Y_{\chi}^{(n)} (T)}{Y_{\chi,{\rm instant.}} } \simeq f(n)
\begin{cases}
\dfrac{8}{5}\left(\dfrac{n+1}{6-n}\right)\,, & n<6\\[15pt]
\dfrac{56}{5}\ln\left(\dfrac{T_{\rm max}}{T_{RH}}\right)\,, & n=6 \\[15pt]
\dfrac{8}{5}\left(\dfrac{n+1}{n-6}\right)\left(\dfrac{T_{\rm max}}{T_{RH}}\right)^{n-6}\,, & n>6
\end{cases} \,,
\eeq
where we have inserted a function $f(n)$ shown in Fig.~\ref{fig:fnfig},
which corrects the analytic result discussed above, with the 
exact numerical evaluation. This correction is necessary as, around $v\sim 1$, the approximation (\ref{Tv}) to the plasma temperature is not accurate, due to the shift of the equation of state parameter from $w\approx 0$ to $w\approx 1/3$; moreover, entropy production continues beyond $v=1$, which further dilutes the dark matter yield below the analytical approximation. Note that, nevertheless, the correction is not large, $0.4 \lesssim f(n) \lesssim 3.3$ for $n> 0$. Eq.~(\ref{Yratio}) is one of the main results of this paper. 

We see from eq.~(\ref{Yratio}) that as $n$ increases, the final result for the abundance is increasingly sensitive to the highest temperature, and the details of reheating are relevant.  In particular, physically different results are obtained for $n<6$ vs. $n\geq 6$, as already noted in 
\cite{Tmax2} (that only focused on the $n<6$ case). For $n<6$ the more accurate result (\ref{Yabove}) corrects the  instantaneous reheating result by a factor of  $\mathcal{O}(1)$. For a steeper dependence of the cross section on the  temperature, the final dark matter abundance can be significantly different from the naive expectation.  In particular, in terms of the inflaton decay rate, the enhancement for the $n=6$ case can be equivalently rewritten as
\beq\label{Yratio2}
R_{\chi} \simeq 
 1.14\ln\left(\dfrac{m_{\phi}}{\Gamma_{\phi}}\right) - 3.17 \,.
\eeq

\section{Representative examples} 

In this section we consider three representative cases characterized by the thermal cross section (\ref{sigmav}), with three different values of the coefficient $n$. These are: (1)  Gravitino production in low scale supersymmetry models (with a single gravitino in the final state). This is characterized by $n=0$; (2) Non-Equilibrium Thermal Dark Matter (NETDM), characterized by $n=2$; (3)  Gravitino production in high scale supersymmetry where production occurs in processes having two gravitinos in final state, leading to $n=6$.

\subsection{Single Gravitino Production}

In commonly studied models of weak scale supersymmetry, in the absence of direct inflaton to gravitino decays,
the dominant scattering source for gravitino production is $X +{\tilde Y} \to {\tilde G} + Z$  or $X + Y \to {\tilde Z} + {\tilde G}$
where $X, Y, Z$ are standard model (SM) particles or their supersymmetric partners. The cross section for the production of the transverse components of the gravitino is simply proportional to $(1/M_P^2)$ \cite{nos,ehnos,kl,ekn,Kawasaki:1994af}.  However, when the mass of the gravitino
is less than the gaugino masses (and in particular the gluino mass), the cross section for the production of the
longitudinal components is enhanced by a factor of $(m_{\tilde g}/m_{3/2})^2$ \cite{mmy,enor,Giudice:1999am,bbb,Pradler:2006qh,ps2,rs,EGNOP}.

The thermally-averaged cross section for the Standard Model 
$SU(3)_c\times SU(2)_L\times U(1)_Y$ gauge group was calculated in~\cite{bbb,Pradler:2006qh,rs}. 
The dominant contributions to the cross section can be parametrized as
\beq
\langle \sigma_{\rm tot}v_{\rm rel}\rangle  =  \langle \sigma_{\rm tot}v_{\rm rel}\rangle_{\rm top}
+ \langle \sigma_{\rm tot}v_{\rm rel}\rangle_{\rm gauge} \,,  
\eeq
with
\beq
\langle \sigma_{\rm tot}v_{\rm rel}\rangle_{\rm top} =  1.29\,\frac{|y_t|^2}{M_P^2}\left[1+\frac{A_t^2}{3m_{3/2}^2}\right] \,,
\eeq
where $A_t$ is the top-quark supersymmetry-breaking trilinear coupling, and
\begin{eqnarray}
\langle \sigma_{\rm tot}v_{\rm rel}\rangle_{\rm gauge} & = & \sum_{i=1}^3 \frac{3\pi c_ig_i^2}{16 \zeta(3) M_P^2} \left[1+\frac{m_{\tilde{g}_i}^2}{3m_{3/2}^2}\right]
\ln\left(\frac{k_i}{g_i}\right) \nonumber \\
& & \!\!\!\!  \!\!\!\!  \!\!\!\!  \!\!\!\!  \!\!\!\! 
=  \frac{26.24}{M_P^2}  \left[\left(1+0.558\,\frac{m_{1/2}^2}{m_{3/2}^2}\right) - 0.011 \left(1+3.062\,\frac{m_{1/2}^2}{m_{3/2}^2}\right) \log\left(\frac{T}{10^{10}\,{\rm GeV}}\right)\right] \, ,
\label{ck}
\end{eqnarray}
where the $m_{\tilde{g}_i}$ are the gaugino masses and the constants $c_i,k_i$ depend on the gauge group, as shown in Table~\ref{table:gauge}. The second line of (\ref{ck}) was obtained in ref.~\cite{EGNOP} from a 
 fit to the result of \cite{rs} using the parametrization of \cite{Pradler:2006qh}, 
under the assumption of a unified gauge coupling $\alpha=1/24$ and universal
gaugino masses $m_{1/2}$ at the scale $M_{\rm GUT} = 2\times 10^{16}\,$GeV (see \cite{EGNOP} for details). 
Note that the first term in the gaugino mass-dependent factors $(1+m_{\tilde{g}_i}^2/3m_{3/2}^2)$ corresponds to the production of the transversally polarized gravitino, while the second term is associated with the production of the longitudinal (Goldstino) component.
For $m_{3/2} \ll m_{\tilde {g_i}}$, the production of the longitudinal components dominates.
\begin{table}[ht]
\centering
\begin{tabular}[c]{c c c c}
\hline \hline Gauge group & $g_i$ & $c_i$ & $k_i$\\ 
\hline $U(1)_Y$ & $g'$ & 9.90 & 1.469\\ 
 $SU(2)_L$ & $g$ & 20.77 & 2.071\\ 
 $SU(3)_c$ & $g_s$ & 43.34 & 3.041\\ 
\hline \hline
\end{tabular} 
\caption{\em The values of the constants $c_i$ and $k_i$ in the parameterization (\protect\ref{ck})
for the Standard Model gauge groups $U(1)_Y$, $SU(2)_L$, and $SU(3)_c$. See \cite{EGNOP} for details.
}
\label{table:gauge}
\end{table}

\begin{figure}[ht!]
\centering
    \scalebox{0.8}{\includegraphics{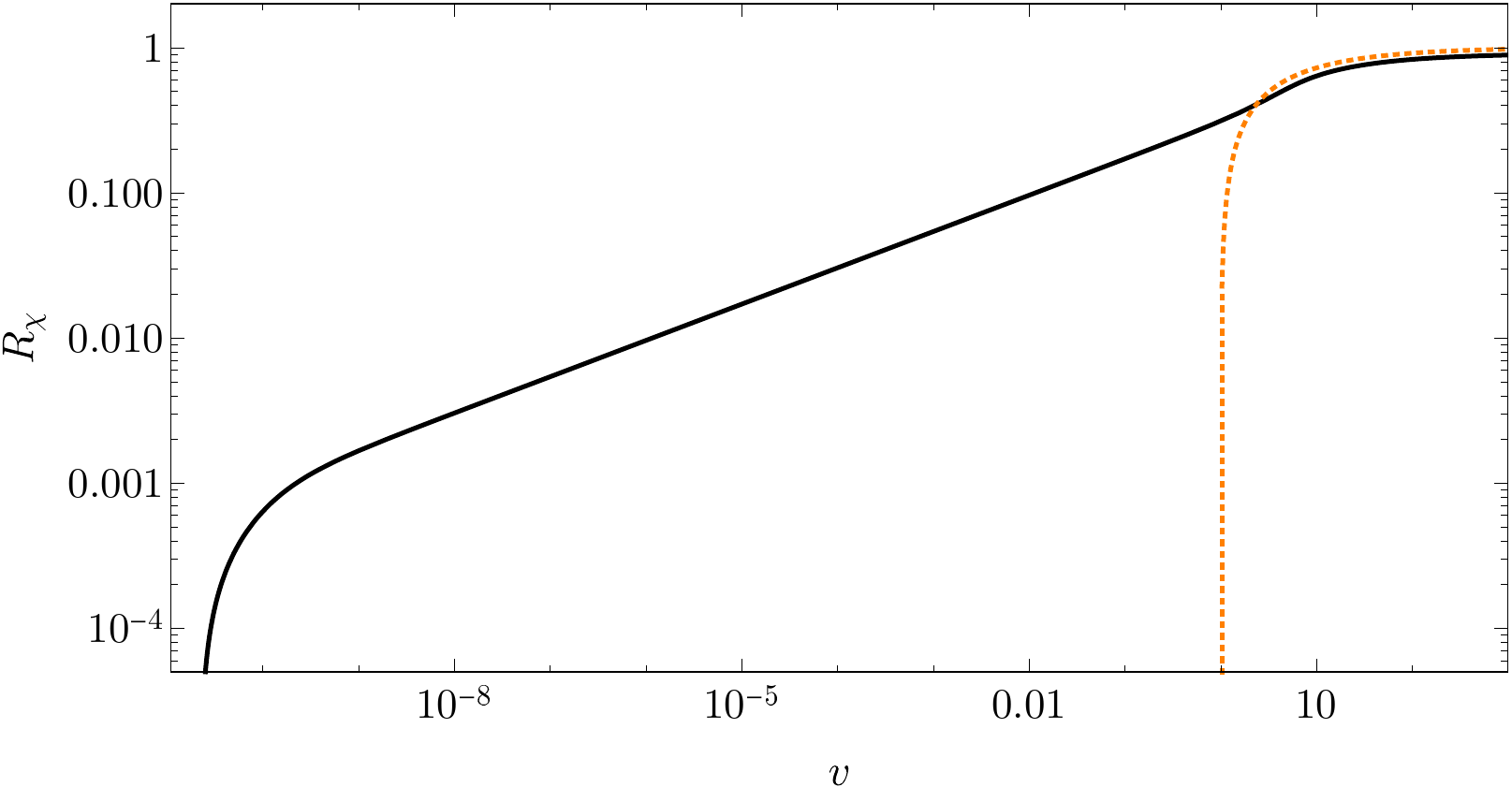}}
    \caption{\em Dark matter yield during and after reheating with $n=0$; here $\Gamma_{\phi}=10^{-11}$ $m_\phi$. The numerical result using $R_\chi^{(n)}$ (eq.~(\ref{Yratio})) with $n=0$ is shown as the continuous black curve. The orange dotted curve is the instantaneous reheating solution from $R_{\chi,{\rm instant.}}(T)$ (eq.~(\ref{Rinst})).
        }
    \label{fig:n0fig}
\end{figure}

Ignoring the logarithmic dependence in eq.~(\ref{ck}), the cross section is constant corresponding to 
$n=0$ in eq.~(\ref{sigmav}). Figure~\ref{fig:n0fig} shows the comparison between the fully numerical calculation (black, continuous), using
$R_\chi^{(n)}$ in eq.~(\ref{Yratio}) with $n=0$,  and the instantaneous reheating result (orange, dotted), given by $R_{\chi,{\rm instant.}}(T)$ from eq.~(\ref{Rinst}). 
The latter by definition asymptotes to 1 at late times (large $v$). 
As it is clear, the instantaneous approximation slightly overestimates the true gravitino abundance by a factor of $\sim 1.1$, as expected from eq.~(\ref{Yratio}).
More importantly we see that gravitino production prior to the end of reheating can be ignored, as
any production between $T_{RH}$ and $T_{\rm max}$ is diluted by the bulk of the entropy produced in later inflaton decays.

\subsection{NETDM Production}
In the standard gravitino production mechanism discussed above, the gravitino is produced from the thermal bath,
but it never itself achieves thermal equilibrium with the bath.  Dark matter particles coupled to the thermal bath through
a heavy mediator (such as an intermediate scale gauge boson) can also be produced from the thermal bath
while never achieving thermal equilibrium.  
Such NETDM candidates \cite{moqz,mnoqz,brian}, may arise in non-supersymmetric
grand unified theories such as SO(10) when a SM singlet component of either a {\bf 45, 54} or {\bf 210} representation of SO(10)
is the dark matter \cite{mnoqz,noz}.  

\begin{figure}[t]
\centering
\includegraphics[width=0.5\columnwidth]{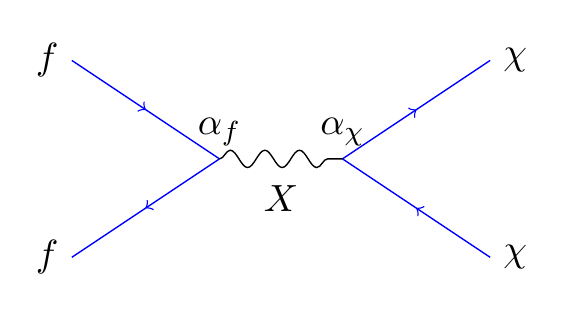}
\caption{
\footnotesize
{
Feynman diagram depicting the freeze in production of the dark matter $\chi$
through a heavy $X$ mediator.
}
}
\label{Fig:feynman}
\end{figure}

Here, we consider the production of a fermionic dark matter candidate, $\chi$, via a  $2\leftrightarrow 2$ process mediated by the exchange of a heavy gauge boson $X$. For definiteness, we assume that the parent SM particles  (denoted by $f$) are also fermions, leading to the  diagram depicted in Fig.~\ref{Fig:feynman} with matrix element squared

\beq
|\mathcal{M}|^2 = \frac{\alpha_{f}^2\alpha_{\chi}^2s^2}{(s-m_X^2)^2}(1+\cos^2\theta)\,.
\eeq
Here $\alpha_{f,\chi}$ denote the gauge couplings, while $\theta$ is the angle between the incoming and outgoing particles in the CM frame. The same amplitude is obtained for a scalar mediator $X$, with $\alpha_{f,\chi}$ playing the role of Yukawa couplings without the $\cos^2 \theta$. The scattering cross section can be obtained in a straightforward way, 
\beq\label{schsg}
\sigma_{\chi\chi\leftrightarrow ff} =  \frac{\alpha_{f}^2\alpha_{\chi}^2 s}{12\pi(s-m_X^2)^2}\,.
\eeq
The dark matter abundance follows eq.~(\ref{set4}), with  $n_{\chi}^{\rm eq}=g_{\chi}n_{\rm rad}$. 
The thermally averaged cross section can be computed in the ultrarelativistic limit $T\gg m_{\chi}$ as \cite{Gondolo:1990dk,moqz}
\beq\label{sigmasimp}
\langle \sigma v \rangle = \frac{49}{18}
\frac{N_f \alpha_f^2 \alpha_\chi^2}{m_X^4 \pi}\left[  \frac{\zeta(4)}{\zeta(3)}\right]^2 T^2
\simeq 2.2~N_f~
\frac{\alpha_f^2 \alpha_\chi^2}{\pi m_X^4}  ~T^2\, ,
\eeq
where $N_f$ the number of SM fermions coupling with the mediator $X$
and we have assumed $T\gg m_{\chi}$ and $T \ll m_X$.
The final expression is of the form
(\ref{sigmav}), with $n=2$, $\lambda=\alpha_{f}^2\alpha_{\chi}^2$ and $M=m_{X}$.  We use this expression in the system of  equations  (\ref{set1})-(\ref{set2})-(\ref{set4}), which we integrate numerically, under the assumption, characteristic of NETDM, that the dark matter abundance is much smaller than the thermal equilibrium value, $n_{\chi} \ll n_{\chi}^{\rm eq}$. For generality we plot $R_{\chi}$, which scales out all model-dependent factors from the dark matter yield during reheating, and we have assumed a constant $g=g_{RH}$.

\begin{figure}[ht!]
\centering
    \scalebox{0.8}{\includegraphics{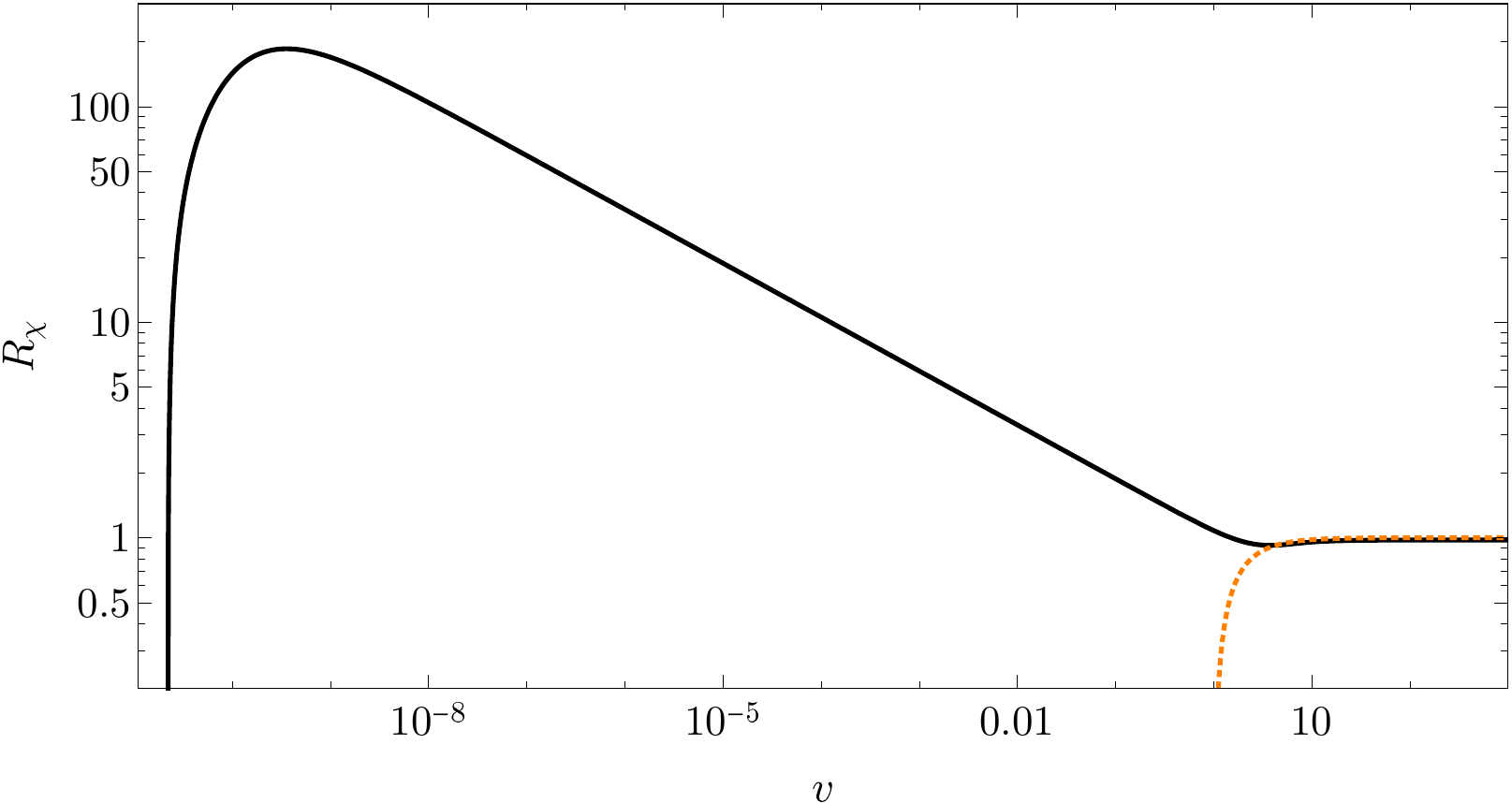}}
    \caption{\em As in Fig.~\ref{fig:n0fig}, for $n=2$.
        }
    \label{fig:n2fig}
\end{figure} 
Figure~\ref{fig:n2fig} shows the comparison between the fully numerical calculation (black, continuous)  using
$R_\chi^{(n)}$ in eq.~(\ref{Yratio}) with $n=2$, and the instantaneous reheating result (orange, dotted)  given by $R_{\chi,{\rm instant.}}(T)$ from eq.~(\ref{Rinst}). It can be seen that the instantaneous approximation minimally overshoots the exact solution by a mere $3\%$, in agreement with (\ref{Yratio}).

\subsection{High Scale Supersymmetry, with two gravitinos final state processes}

Our final example is that of two-gravitino final state processes which are the dominant 
gravitino production mechanisms in high scale supersymmetry models where the only
supersymmetric state below the inflationary scale is the gravitino \cite{bcdm,DMO}.
In this case, the process $X +{\tilde Y} \to {\tilde G} + Z$ is not possible as there are no
supersymmetric particles in the thermal bath and  $X + Y \to {\tilde Z} + {\tilde G}$ is kinematically
forbidden.  Thus, the dominant process becomes  $X + Y \to {\tilde G} + {\tilde G}$ which is highly
suppressed. Since $m_{3/2} \ll m_{\tilde g}$, we expect the cross section to longitudinal modes to dominate
and when accounting for all possible SM initial states, the thermally averaged cross section can be written as \cite{bcdm}
\beq
\langle \sigma v \rangle \simeq 2000 \frac{T^6}{\pi F^4}\,,
\label{Eq:gravitino}
\eeq
where $F = \sqrt{3} M_P m_{3/2}$ is the supersymmetry breaking order parameter.

The strong suppression ($\propto F^4$) of the cross section would indicate that
a relatively high reheating temperature and gravitino mass are required to produce a sufficient quantity of gravitinos to account 
for the observed relic density. 
Indeed for a gravitino mass of 1~EeV, a reheating temperature of approximately 
$5 \times 10^{10}$ GeV is needed \cite{DMO}, placing strong constraints on  inflationary models and supersymmetry breaking
\cite{dgmo}.

\begin{figure}[ht!]
\centering
        \scalebox{0.8}{\includegraphics{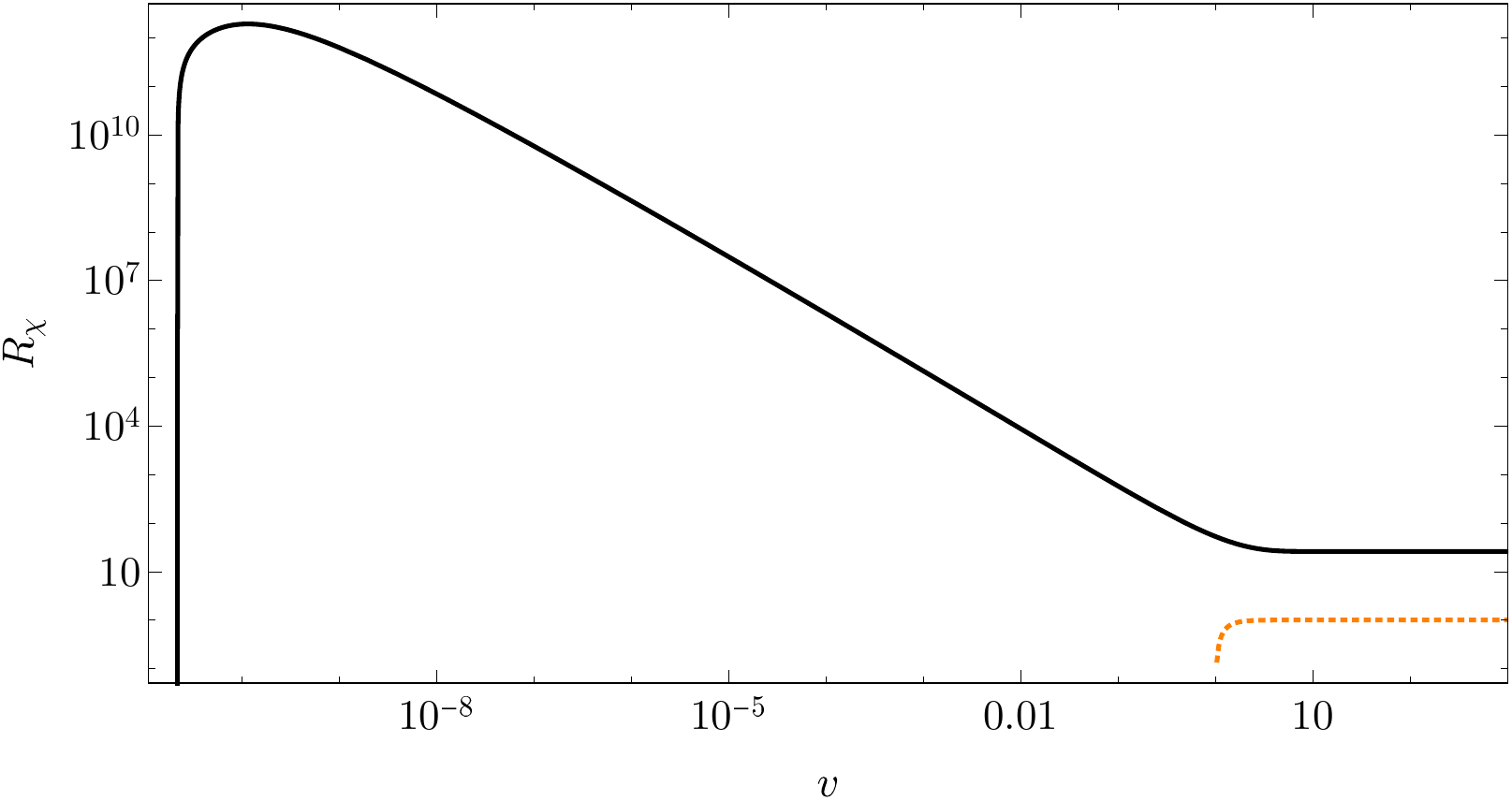}}
    \caption{\em As in Fig.~\ref{fig:n0fig}, for $n=6$.        }
    \label{fig:n6fig}
\end{figure}

Figure~\ref{fig:n6fig} shows the exact and instantaneous results for $R_{\chi}$ in the $n=6$ case. In this case,
one sees that the standard estimate of the dark matter abundance evaluated at 
$T_{RH}$ is not very accurate and the final ratio is $R_{\chi} \sim 25.7$, consistent with the result (\ref{Yratio2}).
From eq.~(\ref{Yinst}) we see that, in order to obtain the correct gravitino dark matter abundance, the reheating temperature should be decreased by a factor $\sim \frac{2}{3}$ with respect to that indicated by the naive assumption of instantaneous decay.

\section{Conclusions}

Reheating after inflation is responsible for the entire matter and radiation
content of the Universe. Thus, understanding the details of this process is crucial to our ability to 
develop models incorporating entropy production, baryogenesis, and dark matter among many other 
important ideas in cosmology.  

In many models of dark matter, including well studied models of supersymmetric dark matter, thermally produced dark matter particles come into thermal equilibrium, and
their final abundance is often determined after they freeze out of the thermal bath.
On the other hand, there are many models 
in which the dark matter candidates never attain thermal 
equilibrium yet are produced from the thermal bath.  Gravitino dark matter is a good example of this situation,
and early estimates of the final gravitino abundance \cite{dg,nos} relied on the instantaneous reheating approximation and the definition of the reheating temperature.  Reheating, however, is not
an instantaneous process, but rather a continuous one. The rapid thermalization of the particles produced in the earliest stages of reheating results in a thermal bath with temperatures potentially much higher than the classically defined reheating temperature.

Here, we have examined the effect of the high temperatures attained during
reheating on the production of dark matter particles. We computed the abundance of a particle produced from the thermal bath with thermally averaged cross section  $\langle \sigma v \rangle \propto T^n$. Eq.~(\ref{Yratio}) provides a  simple result for the discrepancy between the exact abundance, and the naive calculation based on instantaneous reheating. This result can be immediately applied to obtain the exact abundance for a number of particle physics models. 
We considered three specific examples, characterized by three different values of the exponent $n$.  

Two cases, singly produced gravitinos in 
low energy supersymmetric models, and NETDM candidates coupled to the 
SM through heavy mediators, have production 
cross sections with a relatively mild temperature dependence $(n=0$ and $n=2$, respectively). Even in the case of $n=2$, the increased cross section at temperatures $T>T_{RH}$, is not sufficient to 
overcome the dilution from inflaton decays when $T< T_{\rm max}$. However,
we also considered the case of gravitino production in high scale supersymmetric models
when the only non-SM particle lighter than the inflaton is the gravitino. In this case, 
gravitinos must be produced in pairs leading to an additional scale suppression of the cross section which 
in turn, leads to a larger temperature dependence.  Indeed, in this case, $n=6$ and
the production of gravitinos near $T_{\rm max}$ can not be neglected. We found that the true gravitino
abundance exceeds the naive calculation by a factor of $\sim 25$.

\section*{Acknowledgements}

This  work was supported by the France-US PICS no.~06482.
 Y.M.~acknowledges partial support from the European Union’s Horizon 2020 research and
innovation program under the Marie Sklodowska-Curie Grants No.~690575 and No.~674896 and  the ERC advanced grants  
 Higgs@LHC. The work of  K.A.O. and M.P. was supported in part
by DOE grant DE--SC0011842 at the University of Minnesota. The work of M.A.G.G. was partially completed at the Aspen Center for Physics, which is supported by National Science Foundation grant PHY-1066293.

\end{document}